\documentclass[11pt]{article}
\usepackage{latexsym}  
\usepackage{amssymb}
\usepackage{graphicx}
\usepackage{amsmath}

\begin{document}

\begin{center}
{\Large\bf Leptonic $CP$ violating phase in the Yukawaon model} 
  
\vspace{4mm}
{\bf Yoshio Koide$^a$ and Hiroyuki Nishiura$^b$}

${}^a$ {\it Department of Physics, Osaka University, 
Toyonaka, Osaka 560-0043, Japan} \\
{\it E-mail address: koide@kuno-g.phys.sci.osaka-u.ac.jp}

${}^b$ {\it Faculty of Information Science and Technology, 
Osaka Institute of Technology, 
Hirakata, Osaka 573-0196, Japan}\\
{\it E-mail address: hiroyuki.nishiura@oit.ac.jp}

\date{\today}
\end{center}

\vspace{3mm}

\begin{abstract}
In the so-called ``Yukawaon" model, the (effective) Yukawa coupling 
constants $Y_f^{eff}$ are given by vacuum expectation values (VEVs) of 
scalars $Y_f$ (Yukawaons) with $3\times 3$ components. 
In this brief article, we change VEV forms $\langle Y_f \rangle$ in the
previous paper into a unified form. 
Therefore, parameter fitting for quark and lepton masses and mixings 
is revised. 
Especially, we obtain predicted values of neutrino mixing $\sin^2 2\theta_{13}$
and a leptonic $CP$ violating phase $\delta_{CP}^\ell$ that are consistent
with the observed curve in the $(\sin^2 2\theta_{13}, \delta_{CP}^\ell)$ 
reported by T2K group recently.
\end{abstract}

PCAC numbers:  
  11.30.Hv, 
  12.15.Ff, 
  14.60.Pq,  
  12.60.-i, 

\vspace{3mm}

\noindent{\large\bf 1 \ Introduction}

Now, measurement of $CP$ violating phase $\delta_{CP}^\ell$ in the lepton
sector is within our reach because of the recent development of neutrino
physics\cite{neutrino_PDG14}.
The measurement is very important to check quark and lepton mass matrix
models currently proposed.
At the same time, for model builders, it is urgently required to predict 
an explicit value of $\delta_{CP}^\ell$ together with mixing value
$\sin^2 2 \theta_{13}$ based on their models.
So, we estimate a value of $\delta_{CP}^\ell$ based on the so-called 
Yukawaon model \cite{yukawaon_models,K-N_PRD14}, 
which is a unified mass matrix model of quarks and leptons, 
and which is a kind of flavon model \cite{flavon}.
 
In the Yukawaon model, 
the (effective) Yukawa coupling constants $Y_f^{eff}$ are given 
by vacuum expectation values (VEVs) of 
scalars $Y_f$ (Yukawaons) with $({\bf 8}+{\bf 1})$ of U(3) family symmetry:
$$
(Y_f^{eff})_i^{\ j} = \frac{y_f}{\Lambda} \langle Y_f\rangle_i^{\ j} 
\ \ \ \ (f=u, d, \nu, e),
\eqno(1)
$$
where $\Lambda$ is a scale of the effective theory.
In understanding flavor physics from a view of a non-Abelian family symmetry, 
the conventional Yukawa interactions explicitly 
break its family symmetry. 
It is only when the conventional Yukawa coupling constants 
are supposed to be given by Eq.(1) that we can build a model with an unbroken family 
symmetry.

The characteristic point of the Yukawaon model is 
the following point: 
The quark and lepton mass matrices are
described by using only the observed values of charged lepton masses 
$(m_e, m_\mu, m_\tau)$ as input parameters with family-number dependent values;
thereby, we investigate 
whether we can describe all other observed mass spectra 
(quark and neutrino mass spectra) and mixings 
(the Cabibbo-Kobayashi-Maskawa \cite{CKM} (CKM) mixing and 
the Pontecorvo-Maki-Nakagawa-Sakata \cite{PMNS} (PMNS) mixing) without 
using any other family number-dependent parameters. 
Here, terminology ``family number-independent parameters" means, 
for example, coefficients of a unit matrix ${\bf 1}$, 
a democratic matrix $X_3$, and so on, where
$$
{\bf 1} = \left( 
\begin{array}{ccc}
1 & 0 & 0 \\
0 & 1 & 0 \\
0 & 0 & 1 
\end{array} \right) , \ \ \ \ \ 
X_3 = \frac{1}{3} \left( 
\begin{array}{ccc}
1 & 1 & 1 \\
1 & 1 & 1 \\
1 & 1 & 1 
\end{array} \right) . 
\eqno(2)
$$

In the previous paper, the form of $\langle Y_d \rangle$ in the down-quark 
sector has been supposed to be unnaturally different from those in other sectors.
In this paper, we revise the form of $\langle Y_f \rangle$ so that it 
takes a unified form for all sectors as given Eq.(3) in the next section. 
Accordingly, parameter fitting for quark and lepton masses and mixings 
is also revised as given in Secs.3 and 4. 
Especially, it is shown in Sec.4 that we obtain predicted values for neutrino mixing 
$\sin^2 2\theta_{13}$ and a leptonic $CP$ violating phase 
$\delta_{CP}^\ell$ that are consistent with 
the observed curve in the $(\sin^2 2\theta_{13}, \delta_{CP}^\ell)$ plane 
reported by T2K group \cite{T2K_PRL14} recently.


\vspace{3mm}

\noindent{\large\bf 2 \ Models}

Hereafter, for convenience, we use the notation 
$\hat{A}$, $A$ and $\bar{A}$ for fields with ${\bf 8}+{\bf 1}$,
${\bf 6}$ and ${\bf 6}^*$ of U(3), respectively. 
Explicit forms of VEV relations among the Yukawaon in this paper are given by 
$$
\langle \hat{Y}_f \rangle_i^{\ j} = k_f \left[ \langle \Phi_f \rangle_{ik}
 \langle \bar{\Phi}_f \rangle^{kj} + \xi_f {\bf 1}_i^{\ j} \right] \ \ \ 
(f=e,\nu, d, u),
\eqno(3)
$$
$$
 \langle \Phi_f \rangle_{ij} = k'_f  \langle \Phi_0 \rangle_{i\alpha}
  \langle \bar{S}_f \rangle^{\alpha \beta}
  \langle {\Phi}_0^T\rangle_{\beta j} , \ \ \ 
\langle \bar{\Phi}_f \rangle^{ij} = k'_f  \langle \bar{\Phi}_0 \rangle^{i\alpha}
  \langle{S}_f \rangle_{\alpha \beta}
  \langle \bar{\Phi}_0^T\rangle^{\beta j} , \ \ \ 
(f= e, \nu) , 
  \eqno(4)
$$
$$
\langle \bar{E}_u \rangle^{ik}  \langle {\Phi_u} \rangle_{kl} 
\langle \bar{E}_u \rangle^{lj} =
 \langle \bar{\Phi}_0 \rangle^{i\alpha} \langle S_u \rangle_{\alpha\beta} 
\langle \bar{\Phi}_0^T \rangle^{\beta j} , \ \ \ 
\langle {E}_u \rangle_{ik}  \langle \bar{\Phi}_u \rangle^{kl} 
\langle {E}_u \rangle_{lj} =
 \langle {\Phi}_0 \rangle_{i\alpha} \langle \bar{S}_u \rangle^{\alpha\beta} 
\langle {\Phi}_0^T \rangle_{\beta j} , 
\eqno(5)
$$
$$
\langle \bar{P}_d \rangle^{ik}  \langle \Phi_d \rangle_{kl} 
\langle  \bar{P}_d \rangle^{lj} =
\langle  \bar{\Phi}_0\rangle^{i\alpha}  \langle S_d\rangle_{\alpha\beta} 
\langle \bar{\Phi}_0^T\rangle^{\beta j} , \ \ \ 
\langle {P}_d \rangle_{ik}  \langle \bar{\Phi}_d \rangle^{kl} 
\langle  {P}_d \rangle_{lj} =
\langle  {\Phi}_0\rangle_{i\alpha}  \langle \bar{S}_d\rangle^{\alpha\beta} 
\langle {\Phi}_0^T\rangle_{\beta j} , 
\eqno(6)
$$
$$
 \langle S_f \rangle_{\alpha \beta} = ( {\bf 1} +a_f X_3)_{\alpha\beta} ,
 \ \ \ 
  \langle \bar{S}_f \rangle^{\alpha \beta} = ( {\bf 1} +a_f X_3)^{\alpha\beta} ,
\eqno(7)
$$
where $\langle {E}\rangle = {\bf 1}$, and 
indices $\alpha, \beta, \cdots$ are of another family symmetry 
U(3)$'$. 
We consider that the form (7) is due to a symmetry breaking 
U(3)$' \rightarrow$ S$_3$ at $\mu=\Lambda'$.  
The $\xi_f$ terms in Eq.(3) will be discussed later.
Here, the VEV matrices $\hat{Y}_e$, $\hat{Y}_\nu$, $\hat{Y}_u$ 
and $\hat{Y}_d$ correspond to charged lepton mass matrix $M_e$, 
neutrino Dirac mass matrix $M_{Dirac}$, up-quark mass matrix $M_u$, 
and down-quark mass matrix $M_d$, respectively. 
Hereafter, we drop flavor-independent factors in those VEV matrices, 
because we deal with only mass ratios and mixings in this paper.

The VEV structures are essentially the same as the previous paper 
\cite{K-N_PRD14}. 
However, we have done the following minor changes from 
the previous paper:
(i) In the previous paper, 
$\langle \hat{Y}_d \rangle$ and $\langle \Phi_d \rangle$ were given by 
$\langle \hat{Y}_d \rangle = \langle\Phi_d\rangle 
\langle\bar{\Phi}_d \rangle$ and $\langle \Phi_d \rangle= \langle \Phi_0 \rangle
\langle \bar{S}_d \rangle \langle \Phi_0 \rangle + \xi'_d {\bf 1} $, 
respectively, differently from other sectors.
However, it is unnatural that such a term $\xi'_d {\bf 1}$ appears only 
in the VEV of $\Phi_d$.
In this paper, we remove the $\xi'_d {\bf 1}$ term from the $\Phi_d$ 
and unify the appearance place of the ${\bf 1}$ terms that appear in 
$\langle \hat{Y}_f \rangle$ common to all sectors as shown in Eq.(3).
(ii) Along with the changing of the VEV structure in the down-quark sector,
a phase matrix $P_u$ in the previous paper is moved to the down-quark 
sector as shown in Eq.(6).
For convenience, $\bar{E}$ in Eq.(5) and $\bar{P}_d$ in Eq.(6) were 
exchanged with $\bar{P}_u$ and $\bar{E}$ in the previous paper, respectively. 

Neutrino mass matrix $M_\nu$ is given by a seesaw type 
$$
(M_\nu)^{ij} = \langle \hat{Y}_\nu^T \rangle^i_{\ k}
\langle Y_R^{-1}\rangle^{kl} \langle \hat{Y}_\nu \rangle_l^{\ j} ,
\eqno(8) 
$$
as in the previous paper \cite{K-N_PRD14}, 
where
$$
\langle Y_R\rangle_{ij} =  \langle \hat{Y}_e\rangle_i^{\ k}
\langle \Phi_u\rangle_{kj} + \langle \Phi_u\rangle_{ik} 
\langle \hat{Y}_e^T\rangle^k_{\ j} .
\eqno(9)
$$

In general, we can choose either one 
in two cases, (a) $\langle \bar{A} \rangle = \langle {A} \rangle^{*}$ or (b)
$\langle \bar{A} \rangle = \langle {A} \rangle$, for VEV matrices $\langle A\rangle$ 
and $\langle \bar{A} \rangle$ under the $D$-term condition. 
We assume the type (b) for $\Phi_f$ and $S_f$, while the type (a) for $P_d$:
$$
\langle P_d \rangle = v_P {\rm diag}(e^{i\phi_1},\ e^{i\phi_2},\ 1) , \ \ 
\langle \bar{P}_d \rangle = v_P {\rm diag}(e^{-i\phi_1},\ e^{-i\phi_2},\ 1) . \ \
\eqno(10)
$$ 

In order to distinguish each Yukawaon from the others, we assume that
$\hat{Y}_f$ have different $R$ charges from each other together 
with considering 
$R$-charge conservation [a global U(1) symmetry in $N=1$ supersymmetry].
The $R$-charge assignments are essentially not changed from the previous paper
\cite{K-N_PRD14} except for $E_u$ and $P_d$.

Since we consider that the charged lepton mass matrix is the most fundamental  
one, we assume $a_e=0$ and $\xi_e=0$.
Then, $\langle \Phi_0 \rangle$ is expressed as follows:
$$
\langle \Phi_0 \rangle = \langle \bar{\Phi}_0 \rangle \equiv  
{\rm diag} (x_1, x_2, x_3) \propto 
{\rm diag} (m_e^{1/4}, m_\mu^{1/4}, m_\tau^{1/4}),
\eqno(11)
$$
from the $D$-term condition, where $x_i$ are real and
those are normalized as $x_1^2+x_2^2+x_3^2 =1$.

Now, let us give a brief review of the derivation of $\xi_f$ terms.
We assume the following superpotential for $\hat{Y}_f$  ($f=\nu, e, u, d$),  
with introducing flavons $\hat{\Theta}_f$
$$
W_{\hat{Y}} =  \sum_{f=\nu, e, u, d} \left[ \left( \mu_f (\hat{Y}_f)_i^{\ j}   
+ \lambda_{f} ( \Phi_f)_{ik}(\bar{\Phi}_f)^{kj}  \right)
 (\hat{\Theta}_f)_j^{\ i} 
+ \left( \mu'_f (\hat{Y}_f)_i^{\ i}   
+ \lambda'_{f} ( \Phi_f)_{ik}(\bar{\Phi}_f)^{ki}  \right)
 (\hat{\Theta}_f)_j^{\ j} \right].
\eqno(12)
$$
(Here, we have assumed that only $\hat{\Theta}_f$ can be allowed to appear
as a form ${\rm Tr}[\hat{\Theta}]$ in the superpotential.)
Then, a SUSY vacuum condition $\partial W_{\hat{Y}}/\partial \hat{\Theta}_f=0$ 
leads to VEV relation
$$
\langle \hat{Y}_f \rangle = \langle \Phi_f\rangle \langle\bar{\Phi}_f\rangle 
+\xi_f {\bf 1} , 
\eqno(13)
$$
where
$$
\xi_f = -\frac{\mu'_f}{\mu_f} \left( {\rm Tr} [\langle \hat{Y}_f \rangle] 
+\frac{\lambda'_f}{\mu'_f}{\rm Tr}[\langle  \Phi_f\rangle 
\langle \bar{\Phi}_f\rangle]\right) = 
- \frac{\lambda_f/\mu_f -\lambda'_f/\mu'_f}{1 -3\mu'_f/\mu_f} 
{\rm Tr}[\langle  \Phi_f\rangle \langle  \bar{\Phi}_f\rangle] .
\eqno(14)
$$
Here we have assumed that all VEVs of flavons $\hat{\Theta}$
take $\langle\hat{\Theta} \rangle= 0$, so that SUSY vacuum conditions 
for other flavons do not bring any additional VEV relations.
As seen in Eq.(14), if $\langle \Phi_f \rangle$ is complex, 
then the coefficient $\xi_f$ becomes complex too. 
Although the derivation discussed above was given in the previous work \cite{K-N_PRD14},
we considered that the effect of the phase of $\xi_\nu$ is negligibly small, 
so that we treated $\xi_\nu$ as a real parameter approximately in the previous work.
However, in this paper, we found that the phase of $\xi_\nu$ affects not 
a little on our parameter fitting. 

\vspace{2mm}

\noindent{\large\bf 3 \ Parameter fitting}

{\it General:}
We summarize our mass matrices $M_f$ ($\langle Y_f \rangle$) as follows:
$$
Y_e  = \Phi_e \bar{\Phi}_e +\xi_e {\bf 1}, \ \ \ \ 
\Phi_e =\bar{\Phi}_e =\Phi_0 ( {\bf 1} + 
a_e X_3 )\Phi_0 , \ \ \ \ (a_e=0, \, \xi_e=0),
\eqno(15)
$$
$$
Y_\nu = \Phi_\nu \bar{\Phi}_\nu +\xi_\nu e^{i\beta_\nu}{\bf 1}, \ \ \ 
\Phi_\nu =\bar{\Phi}_\nu =\Phi_0 ( {\bf 1} + a_\nu e^{i\alpha_\nu} X_3) \Phi_0 ,
\eqno(16) 
$$
$$
Y_u = \Phi_u \bar{\Phi}_u  +\xi_u {\bf 1} , \ \ \ \ 
\Phi_u = \bar{\Phi}_u = \Phi_0  \left( {\bf 1} + a_u  X_3 \right) \Phi_0, 
 \eqno(17)
$$
$$
\begin{array}{ll}
Y_d = \Phi_d \bar{\Phi}_d + \xi_d e^{i\beta_d} {\bf 1},  &
 \Phi_d = P_d^*  \Phi_0  \left( {\bf 1} + 
a_d  e^{i\alpha_d}  X_3  \right)  \Phi_0  P_d^* , \\
  & \bar{\Phi}_d = P_d  \Phi_0  \left( {\bf 1} + 
a_d  e^{i\alpha_d}  X_3  \right)  \Phi_0  P_d ,
\end{array}
\eqno(18)
$$
$$
M_\nu = Y_\nu Y_R^{-1} Y_\nu , \ \ \ \ 
Y_R = Y_e \Phi_u  + \Phi_u Y_e . 
\eqno(19)
$$
Here, for convenience, we have dropped 
the notations ``$\langle$", ``$\rangle$" and ``$\hat{\ }$". 
Since we are interested only in the mass ratios and mixings, 
we use dimensionless expressions
$\Phi_0 = {\rm diag}(x_1, x_2, x_3)$ (with $x_1^2+x_2^2+x_3^2=1$),  
$P_d= {\rm diag} (e^{i\phi_1}, e^{i\phi_2},1)$, 
and $E={\bf 1}={\rm diag}(1,1,1)$. 
Therefore, the parameters $a_e$, $a_\nu$, $\cdots$ are redefined 
by Eqs.(15)--(19). 

Since the parameters $a_f$ in Eq.(7) can be complex in general,
we denote $a_f$ as $a_f e^{i \alpha_f}$ by real parameters 
$(a_f, \alpha_f)$. 
The VEV structure of $Y_u$ in the present paper is practically 
unchanged from the previous paper \cite{K-N_PRD14}, so that we inherit
the numerical results in the up-quark sector in the previous work by assuming $\alpha_u=0$.
Since we choose $\alpha_\nu$ and $\alpha_d$ as $\alpha_\nu \neq 0$ 
and $\alpha_d \neq 0$,  we have $\beta_\nu \neq 0$ and $\beta_d \neq 0$ according to Eq.(14).
We have denoted  $\xi_\nu$ and $\xi_d$ in Eq.(3) as $\xi_\nu e^{i\beta_\nu}$ 
 and $\xi_d e^{i\beta_d}$, respectively, in Eqs.(16) and (18). 
 Of course, the parameters $\beta_f$ are fixed by the values
 $(a_f, \alpha_f)$, so that $\beta_f$ are not free parameters.

The explicit values of the parameters $(x_1, x_2, x_3)$ are fixed by Eq.(11) 
as 
$$
(x_1, x_2, x_3) =(0.115144, 0.438873, 0.891141), 
\eqno(20)
$$
where we have normalized $x_i$ as $x_1^2+x_2^2+x_3^2=1$.
Therefore, in the present model, except for the parameters 
$(x_1, x_2, x_3)$, we have ten adjustable parameters,  
$(a_\nu, \alpha_\nu, \xi_\nu)$, $(a_u, \xi_u)$, $(a_d, \alpha_d, \xi_d)$,
and $(\phi_1, \phi_2)$   
for the 16 observable quantities (six mass ratios in the
up-quark, down-quark, and neutrino sectors, four CKM 
mixing parameters, and 4+2 PMNS mixing parameters). 

{\it Quark mass ratios}: 
First, we fix the parameter values $(a_u, \xi_u)$ from the 
observed up-quark mass ratios  \cite{q-mass}
$r^u_{12} \equiv ({m_u}/{m_c})^{1/2} 
= 0.045^{+0.013}_{-0.010}$ and $r^u_{23} \equiv 
({m_c}/{m_t})^{1/2} =0.060 \pm 0.005$ 
at $\mu=m_Z$ \cite{q-mass} as follows, 
$$
(a_u, \xi_u) = (- 1.4715, -0.001521) .
\eqno(21)
$$
Of course, we obtain the same values as those in the previous paper.

Next, we try to fix the parameters $(a_d, \alpha_d, \xi_d)$ in 
the down-quark sector by using input parameters \cite{q-mass}
$r^d_{12} \equiv {m_d}/{m_s} = 0.053^{+0.005}_{-0.003}$ and  
$r^d_{23} \equiv {m_s}/{m_b} = 0.019 \pm 0.006$.
However, since we have three parameters for two input values
$m_d/m_s$ and $m_s/m_b$, we cannot fix our three parameters.
It is more embarrassing that there is no solution of 
$m_s/m_b \sim 0.019$ in the $(a_d, \alpha_d, \xi_d)$ parameter region. 
Nevertheless, we found that the minimal value of $m_s/m_b$ is $m_s/m_b\sim 0.03$ at 
 $(a_d, \alpha_d, \xi_d)\sim (-1.5, 16^\circ, 0.004)$ which
can give a reasonable value of $m_d/m_s$ at the same time too.
Therefore, we take the following values:
$$
(a_d, \alpha_d, \xi_d)=(-1.4735, 15.7^\circ, 0.00400),
\eqno(22)
$$
which leads to predictions
$r_{12}^d= 0.0597$ and $r_{23}^d = 0.0312$. 
Note that the value $r_{23}^d = 0.0312$ is considerably
large compared with $r_{23}^d \simeq 0.019$ by Xing {\it et al.}
\cite{q-mass}, 
while the value is consistent with  $r_{23}^d \simeq 0.031$ by 
Fusaoka and Koide \cite{q-mass_FK}.
The values $m_d(\mu)$ and $m_s(\mu)$ are estimated at a lower energy scale, 
$\mu \sim 1$ GeV, so that we consider that the ratio $r_{12}^d$ 
at $\mu = M_Z$ is reliable. 
On the other hand, the value $m_b(\mu)$ is extracted at a different 
energy scale $\mu \sim 4$ GeV from $\mu \sim 1$ GeV, so that 
the value $m_b(M_Z)$ is affected by the prescription of threshold effects 
at $\mu=m_t$, while the value $m_s(M_Z)$ affected by those at $\mu=m_c$,  
$\mu=m_b$ and $\mu = m_t$.
We consider that as for the ratio $r_{23}^d$ at $\mu = M_Z$ 
the value is still controversial.
Anyhow, we have fixed three parameters $(a_d, \alpha_d, \xi_d)$ only
from two values $m_d/m_s$ and $m_s/m_b$.

{\it CKM mixing}: 
The purpose of the present paper is to discuss PMNS parameters,
especially $CP$ violating phase $\delta^\ell_{CP}$.
However, since our model is to give unified description 
of quarks and leptons, for reference, we give results of CKM 
parameter fitting, too.

Since the parameters $(a_u, \xi_u)$ and $(a_d, \alpha_d, \xi_d)$
have been fixed by the observed quark mass ratios, 
the CKM mixing matrix elements  $|V_{us}|$, 
$|V_{cb}|$, $|V_{ub}|$, and  $|V_{td}|$ are functions of 
the remaining two parameters 
$\phi_1$ and $\phi_2$ defined by Eq.(10).
We use the observed CKM mixing matrix elements \cite{PDG14}  
$|V_{us}|=0.2254 \pm 0.0006$, $|V_{cb}|=0.0414 \pm 0.0012$, 
$|V_{ub}|=0.00355 \pm 0.00015$, and 
$|V_{td}|=0.00886^{+0.00033}_{-0.00032}$. 
(Two of those are used as input values in the present analysis, 
and the remaining two are our predictions as references.)
All the experimental CKM parameters are satisfied by 
fine-tuning the parameters $\phi_1$ and $\phi_2$ as 
$$
(\phi_1, \phi_2)=(-42.0^\circ, -15.1^\circ ), 
\eqno(23)
$$
which leads to the numerical results as follows: $|V_{us}|=0.2255$, $|V_{cb}|=0.0429$, 
$|V_{ub}|=0.00359$, and $|V_{td}|=0.00928$ with $\delta_{CP}^\ell=73.0^\circ$. 
In spite of our aim described in Sec.~1, we are forced to 
introduce family number-dependent parameters $(\phi_1, \phi_2)$ 
in the present model, too,  
as the same as in the previous model \cite{K-N_PRD14}. 
Model building without using parameter $(\phi_1, \phi_2)$ is left to our future task. 

 \vspace{3mm}

\noindent{\large\bf 4 \ Parameter fitting in the PMNS mixing and $CP$ violating
phase $\delta_{CP}^\ell$} 

We have already fixed our seven parameters as Eqs.~(21)--(23).
The remaining free parameters are only $(a_\nu, \alpha_\nu, \xi_\nu)$
in the Dirac neutrino sector.
We determine the parameter values of $(a_\nu, \alpha_\nu, \xi_\nu)$ 
as follows:
$$
(a_\nu, \alpha_\nu, \xi_\nu) = ( -3.54, -18.0^\circ, -0.0238),
\eqno(24)
$$
which are obtained 
so as to reproduce the observed values \cite{PDG14} of the following 
PMNS mixing angles and $R_{\nu}$,
$$
\sin^2 2\theta_{12} = 0.846 \pm 0.021, \ \ \ 
\sin^2 2\theta_{13} = 0.093 \pm 0.008, \\
\eqno(25)
$$
$$
R_{\nu} \equiv \frac{\Delta m_{21}^2}{\Delta m_{32}^2}
=\frac{m_{\nu2}^2 -m_{\nu1}^2}{m_{\nu3}^2 -m_{\nu2}^2}
=\frac{(7.53\pm 0.18) \times 10^{-5}\ {\rm eV}^2}{
(2.44 \pm 0.06) \times 10^{-3}\ {\rm eV}^2} = 
(3.09 \pm 0.15) \times 10^{-2} .
\eqno(26)
$$
We show the $a_\nu$ and $\alpha_\nu$ dependences of the PMNS mixing parameters 
$\sin^2 2\theta_{12}$, $\sin^2 2\theta_{23}$, $\sin^2 2\theta_{13}$, 
and $R_{\nu}$ in Fig.~1(a) and Fig.~1(b), respectively. 
It is found that $R_{\nu}$ is very sensitive to $a_\nu$. 


\begin{figure}[ht]
\begin{picture}(200,200)(0,0)
\includegraphics[height=.3\textheight]{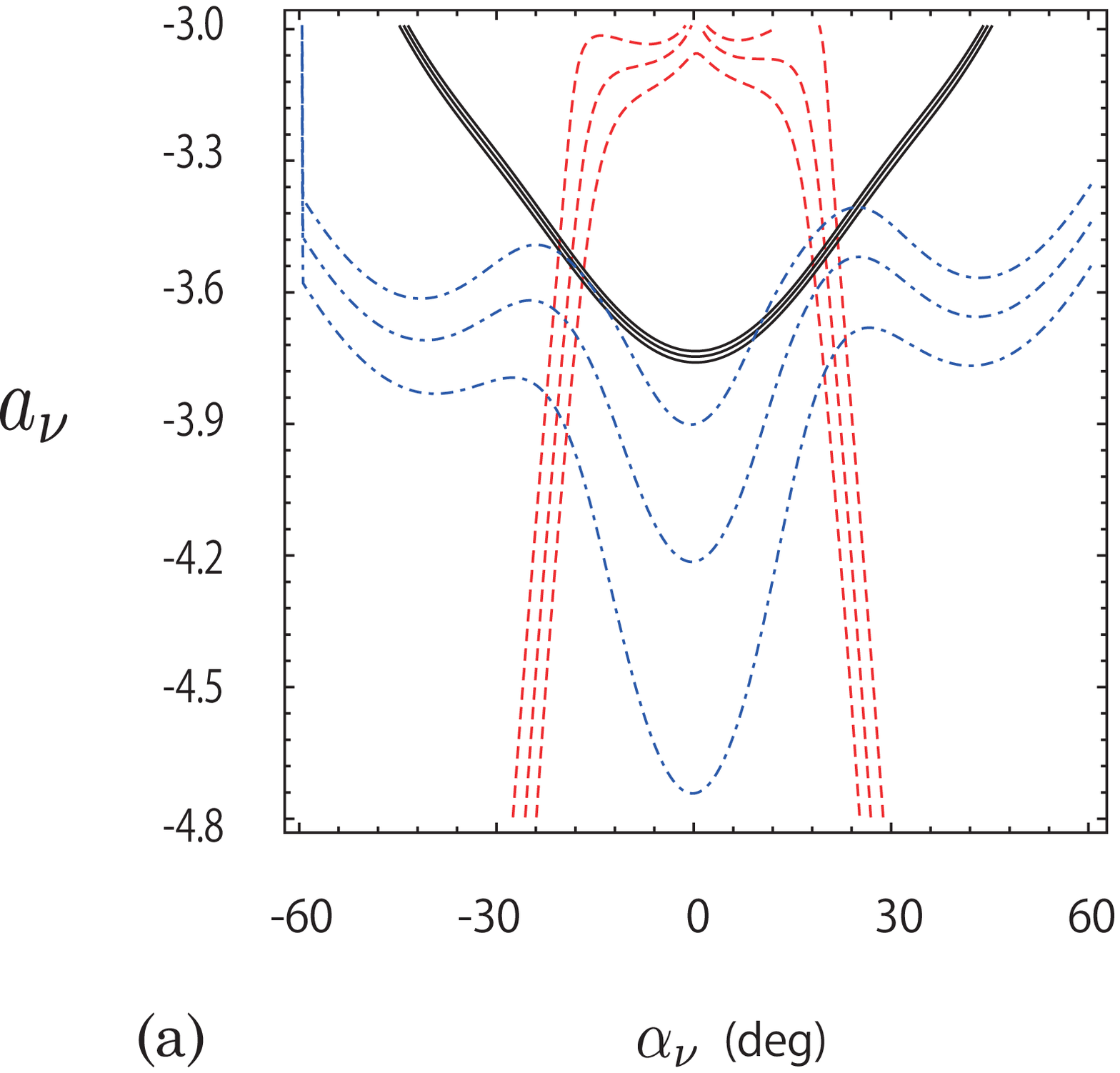}\quad \quad
\includegraphics[height=.3\textheight]{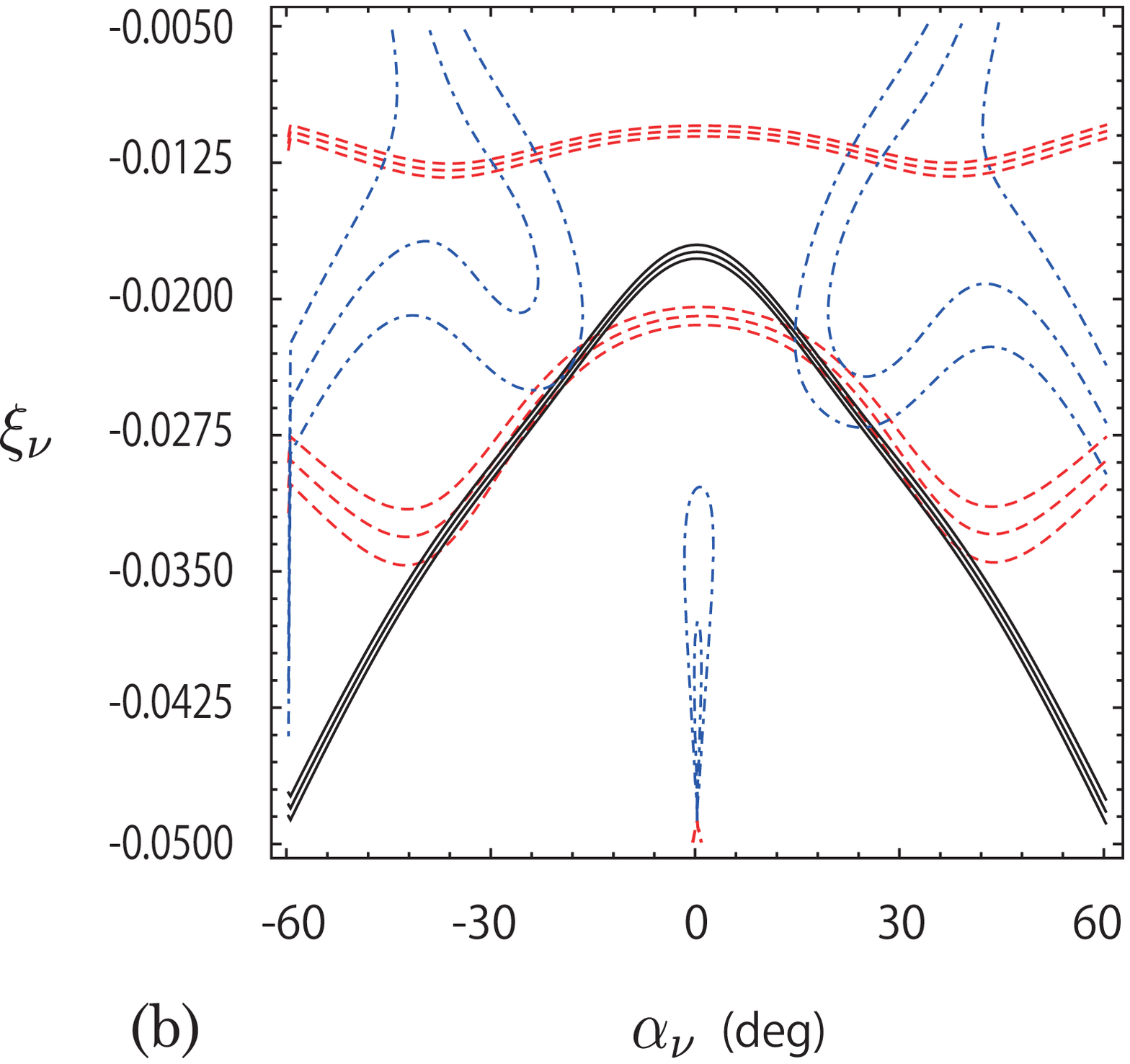}
\end{picture}  
 \caption{Contour curves of the observed center, upper, and lower values of 
the lepton mixing parameters 
$\sin^2 2\theta_{12}$(dashed), $\sin^2 2\theta_{13}$(dot dashed), and the neutrino 
mass squared difference ratio $R_\nu$(solid).  
(a): We draw the curves in the ($\alpha_\nu$, $a_\nu$) plane by taking $\xi_\nu=-0.0238$.
(b): We draw the curves in the ($\alpha_\nu$, $\xi_\nu$) plane by taking $a_\nu=-3.54$. 
}
\label{fig1}
\end{figure}

\begin{figure}[ht]
\begin{picture}(200,200)(0,0)
\includegraphics[height=.3\textheight]{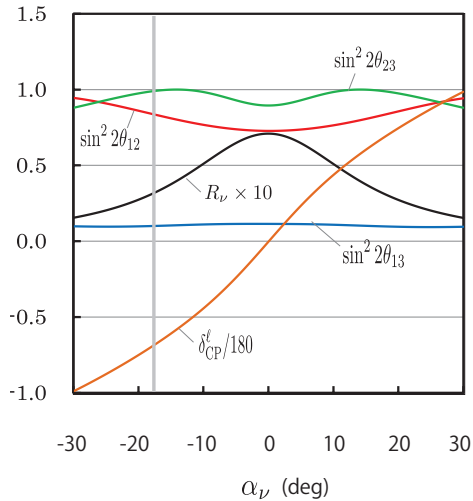}
\end{picture}  
 \caption{ $\alpha_\nu$ dependence of the lepton mixing parameters 
$\sin^2 2\theta_{12}$, 
$\sin^2 2\theta_{23}$, $\sin^2 2\theta_{13}$, $R_\nu$, and 
the leptonic $CP$ violating phase $\delta^\ell_{\rm CP}$. 
We draw the curves of those as functions of $\alpha_\nu$   
for the case of $\xi_\nu=-0.0238$ by taking $a_\nu=-3.54$ (solid). 
}
\label{fig2}
\end{figure}

As seen in Fig.2, we obtain two solutions, 
which are consistent with the  neutrino data except for the data of $\delta_{CP}^\ell$.
However, as seen the best fit curve on the $(\sin^2 2\theta_{13}, \delta_{CP}^\ell)$
plane in Fig.5 in the resent T2K article \cite{T2K_PRL14}, the solution with 
$0 < \delta_{CP}^\ell <\pi$ is obviously ruled out.
Therefore, we adopt the solution with $-\pi <\delta_{CP}^\ell <0$ in our model.
Then, we obtain the predictions of our model
$$
R_\nu = 0.0310, \ \ \sin^2 2\theta_{12}=0.837, \ \ \sin^2 2\theta_{23}=0.988, \ \ 
\sin^2 2\theta_{13}=0.0987, \ \  
\delta_{CP}^\ell =-125^\circ .
\eqno(27)
$$
 
We can  predict neutrino masses, for the parameters given by (21) and (24), 
as follows
$$
m_{\nu 1} \simeq 0.00037\ {\rm eV}, \ \ m_{\nu 2} \simeq 0.00868 \ {\rm eV}, 
\ \ m_{\nu 3} \simeq 0.0501 \ {\rm eV}  ,
\eqno(28)
$$
by using the input value \cite{PDG14}
$\Delta m^2_{32}\simeq 0.00244$ eV$^2$.
We also predict the effective Majorana neutrino mass \cite{Doi1981} 
$\langle m \rangle$ 
in the neutrinoless double beta decay as
$$
\langle m \rangle =\left|m_{\nu 1} (U_{e1})^2 +m_{\nu 2} 
(U_{e2})^2 +m_{\nu 3} (U_{e3})^2\right| 
\simeq 6.0 \times 10^{-3}\ {\rm eV}.
\eqno(29)
$$

Our model predicts $\delta_{CP}^{\ell}= -125^\circ$ for the Dirac $CP$ 
violating phase in the lepton sector, which indicates relatively large  $CP$ 
violating effect in the lepton sector.

\vspace{3mm}

\noindent{\large\bf 5 \ Concluding remarks}

We have tried to describe quark and lepton mass matrices
by using only the observed values of charged lepton masses 
$(m_e, m_\mu, m_\tau)$ as input parameters with family number-dependent values,
except for $P_d$ defined by Eq.(10). 
Thereby, we have investigated  
whether we can describe all other observed mass spectra 
(quark and neutrino mass spectra) and mixings (CKM and 
PMNS mixings) without using any other family number-dependent parameters. 
In conclusion, we have obtained reasonable 
results. We have predicted   
the $CP$ violating phase in the lepton sector as  
 $\delta_{CP}^{\ell} \simeq -125^\circ$ and
$ \sin^2 2 \theta_{13} \simeq 0.099$ in Eq.(27), are consistent with 
the observed curve in the $(\sin^2 2 \theta_{13}, \delta_{CP}^\ell)$ plane 
that has been reported by T2K group  \cite{T2K_PRL14}.
(The predicted value of $\delta_{CP}^\ell$ in the previous paper was
$\delta_{CP}^\ell = - 26^\circ$.)

The origin of the $CP$ violation in the lepton sector is in the phase factor
 $\alpha_\nu$ in the Dirac neutrino mass matrix (16). 
Note that we have taken $\alpha_f =0$ ($f=e, u$) for economy of 
the parameters.  
However, we have been obliged to accept $\alpha_\nu \neq 0$ 
in order to fit the observed value of $\sin^2 2\theta_{13}$.

Although the present model is a minor improved version of the previous
paper \cite{K-N_PRD14}, the predicted value of $\delta_{CP}^\ell$ 
has been changed into a more detectable value in near future 
neutrino observations, and it is consistent with the recent 
T2K result \cite{T2K_PRL14}. 
We expect that the value of $\delta_{CP}^\ell$ will be confirmed 
by near future observations.

\vspace{3mm}

\vspace{2mm}
%

\end{document}